\documentclass[journal]{IEEEtran}
\usepackage{xcolor}

\ifCLASSINFOpdf
   \usepackage[pdftex]{graphicx}
   \DeclareGraphicsExtensions{.pdf,.jpeg,.png}
   \usepackage{caption}
    \usepackage{subcaption}
\else
\fi

\usepackage{amsmath}

\usepackage{multirow}

\usepackage{tikz}
\usepackage{textcomp}
\usepackage{hyperref}
\usepackage{lipsum}
\begin{document}
%
\title{SVSNet: An End-to-end Speaker Voice Similarity Assessment Model}
%
%
%


%
 \author{Cheng-Hung Hu, Yu-Huai Peng, Junichi Yamagishi, Yu Tsao, Hsin-Min Wang }
%

%



\newcommand\copyrighttext{%
  \footnotesize \textcopyright 2022 IEEE. Personal use of this material is permitted.
  Permission from IEEE must be obtained for all other uses, in any current or future
  media, including reprinting/republishing this material for advertising or promotional
  purposes, creating new collective works, for resale or redistribution to servers or
  lists, or reuse of any copyrighted component of this work in other works.}
\newcommand\copyrightnotice{%
\begin{tikzpicture}[remember picture,overlay]
\node[anchor=south,yshift=5pt] at (current page.south) {\fbox{\parbox{\dimexpr\textwidth-\fboxsep-\fboxrule\relax}{\copyrighttext}}};
\end{tikzpicture}%
}

\maketitle
\copyrightnotice

    \vspace*{-0.35cm}
\begin{abstract}
Neural evaluation metrics derived for numerous speech generation tasks have recently attracted great attention. In this paper, we propose SVSNet, the first end-to-end neural network model to assess the speaker voice similarity between converted speech and natural speech for voice conversion tasks. Unlike most neural evaluation metrics that use hand-crafted features, SVSNet directly takes the raw waveform as input to more completely utilize speech information for prediction. SVSNet consists of encoder, co-attention, distance calculation, and prediction modules and is trained in an end-to-end manner. The experimental results on the Voice Conversion Challenge 2018 and 2020 (VCC2018 and VCC2020) datasets show that SVSNet outperforms well-known baseline systems in the assessment of speaker similarity at the utterance and system levels.

\end{abstract}


%
\IEEEpeerreviewmaketitle

\section{Introduction}
%
%
%
%
\IEEEPARstart
The speech generated in voice conversion (VC) tasks remains challenging to effectively evaluate.
In most studies, both objective and subjective evaluation results are reported to compare the performance of VC systems.
For objective evaluation \cite{Das2020}, measurements borrowed from the speaker recognition task are usually used.
For subjective evaluation, a listening test is usually conducted.
Compared with objective evaluation, subjective evaluation incurs more time and cost.
Moreover, to reach unbiased results, a large amount of subjective tests must be carried out \cite{wester2015we}.
However, since the target users of VC are humans, the subjective evaluation results are more important than the objective counterparts.
In our previous study \cite{lo2019mosnet}, we proposed MOSNet, which can predict the mean opinion score (MOS) of human subjective ratings of speech quality and naturalness. 
MOSNet is formed by a Convolutional Neural Network-Bidirectional Long Short-Term Memory (CNN-BLSTM) architecture. 
The results of large-scale human evaluation of Voice Conversion Challenge 2018 (VCC2018) demonstrated that MOSNet achieves a high correlation with human MOS ratings at the system level and a fair correlation at the utterance level. 
 
In \cite{lo2019mosnet}, we also slightly modified MOSNet to predict the similarity scores. The
preliminary results showed that the predicted similarity scores were fairly correlated with human similarity ratings. In this work, to further improve the similarity prediction, we propose a novel assessment model called SVSNet, which has two features:
(1) To more accurately characterize speech signals, SVSNet directly takes the speech waveform as input. 
(2) SVSNet adopts a co-attention mechanism to deal with length mismatch, content mismatch, and switching of paired utterances.
For (1), although hand-crafted features are widely used in many speech processing tasks, such as speaker verification (SV) \cite{snyder2018x, variani2014deep}, VC \cite{hsu2017voice, hsu2016voice}, speech synthesis \cite{li2019neural, shen2018natural}, and speech enhancement \cite{pandey2019new}, we believe that the raw waveform contains the most complete information for two reasons. First, for most hand-crafted features, the phase information is ignored. However, many studies have shown that phase can provide useful information \cite{loweimi2015source, loweimi2017robust, loweimi2018robust}. Second, to compute hand-crafted features, prior knowledge is required about feature extraction specifications, such as window size, shift length, and feature dimension. Improper specifications can lead to ineffective features, which can result in poor prediction performance.
For (2), we want to deal with asymmetry. First, the two input utterances may have different lengths and speech content. Second, when the input order is switched, SVSNet should output the same prediction score. Therefore, we design a special co-attention mechanism, which aligns the two input utterances in both directions. Compared to simple alignment methods such as dynamic time warping (DTW), attention can handle content mismatches. We also argue that co-attention can provide more information for similarity assessment than single-sided attention.
Our experimental results on the VCC2018 \cite{lorenzo2018voice} and VCC2020 \cite{vcc2020} datasets show that SVSNet can predict the similarity score of a VC system quite accurately. As per our knowledge, this is the first deep learning-based model for similarity assessment for VC tasks.

\section{Related Works}

\subsection{Neural Evaluation Metrics}
Conventional evaluation metrics are generally derived on the basis of signal processing and human auditory theories. For example, perceptual evaluation of speech quality (PESQ) \cite{rix2001perceptual} and short-time objective intelligibility (STOI) \cite{taal2010short} are commonly used to evaluate the quality and intelligibility of processed speech. 
The normalized covariance measure (NCM) \cite{ma2009objective} and its extensions \cite{chen2010analysis, chen2012contributions} have been shown to be effective in measuring the intelligibility of normal speech and vocoded speech. 
In addition, some parametric distances are often used to measure the difference between paired voices, such as speech distortion index (SDI) \cite{sdi}, mel-cepstral
distance (MCD) \cite{mcd}, cepstrum distance (Cep) \cite{cep}, segmental signal-to-noise ratio (SSNR) improvement \cite{ssnr}, and scale-invariant source-to-noise ratio (SI-SNR) \cite{luo2019conv}. 
Several studies have indicated that these objective evaluation metrics may not truly reflect human perception \cite{mcd}. Therefore, subjective listening evaluations are usually reported in speech generation studies. Unbiased subjective results, however, require a large number of tests, covering a wide range of listeners (gender, age, and hearing ability) and test samples, which makes listening tests challenging in terms of time and cost.

To address the above issues, several neural evaluation metrics have been proposed. 
For speech enhancement tasks, Quality-Net \cite{qualitynet}, DNSMOS \cite{dnsmos}, 
and STOI-Net \cite{stoinet} were proposed as non-intrusive tools for measuring speech quality and intelligibility.
For VC, MOSNet \cite{lo2019mosnet} and MBNet \cite{leng2021mbnet} were proposed to measure the naturalness of converted speech.
Mittag and Möller \cite{mittag2020deep} proposed an assessment model for the text-to-speech synthesis task. To the best of our knowledge, no prior work has previously established neural evaluation metrics for the similarity assessment of VC tasks. 



    \vspace*{-0.15cm}
\subsection{Similarity Prediction}
The similarity prediction task resembles an SV task, which aims to determine whether the input speech is pronounced by a claimed speaker. For most SV systems, the test utterance and the enrollment utterance are first converted into embedding vectors through a neural network (NN), and then a similarity score between the two embedding vectors is calculated based on a distance function, such as cosine distance or another NN model \cite{snyder2018x,zeng2021attention,zhang2019seq2seq}. 
The organizer of VCC2020 reported the results of speaker similarity evaluation using x-vector \cite{Das2020}. In \cite{saito2021perceptual}, several deep speaker representation learning methods that considered the perceptual similarity among speakers were proposed for multispeaker generative modeling. Experimental results show that compared with speaker-classification-based speaker representation learning, perceptual-similarity-aware speaker representation learning has better performance in several speech generative tasks. The speaker embedding learned in this way may be more suitable for speaker similarity evaluation, but the $N \times N$ speaker-pair similarity matrix of $N$ speakers as the training target is not available for the VC task.

    \vspace*{-0.15cm}
\subsection{Waveform Modeling}
Recently, several approaches have been proposed to incorporate waveform modeling into speech processing tasks, such as speech recognition \cite{parcollet2020e2e}, speech enhancement \cite{pascual2017segan, shifas2020non}, speech separation \cite{zeghidour2020wavesplit,luo2019conv}, speech vocoding \cite{ yamamoto2020parallel,wavenet,kumar2019melgan}, and SV \cite{jung2019rawnet,jung2018complete}. The main idea of these waveform modeling methods is that traditional hand-crafted feature extraction techniques can be substituted by NN models in a data-driven manner.
To effectively model speech waveforms, a dilated architecture has been proposed to increase the
reception field with the same number of model parameters \cite{wavenet}. 
Meanwhile, SincNet \cite{ravanelli2018speaker} processes the raw waveform with a set of parameterizable band-pass filters, where only the low and high cutoff frequencies of the band-pass filters are the parameters to be learned. Learning data-dependent and task-dependent filters provides greater flexibility than fixed feature processing procedures. The effectiveness of SincNet has been demonstrated in several studies \cite{ravanelli2018speaker,  liu2020multichannel,ravanelli2019pytorch}.

    \vspace*{-0.15cm}
\section{Proposed SVSNet}
\begin{figure}
    \centering
    \begin{subfigure}[b]{0.45\textwidth}
        \centering
        \includegraphics[width=1\textwidth]{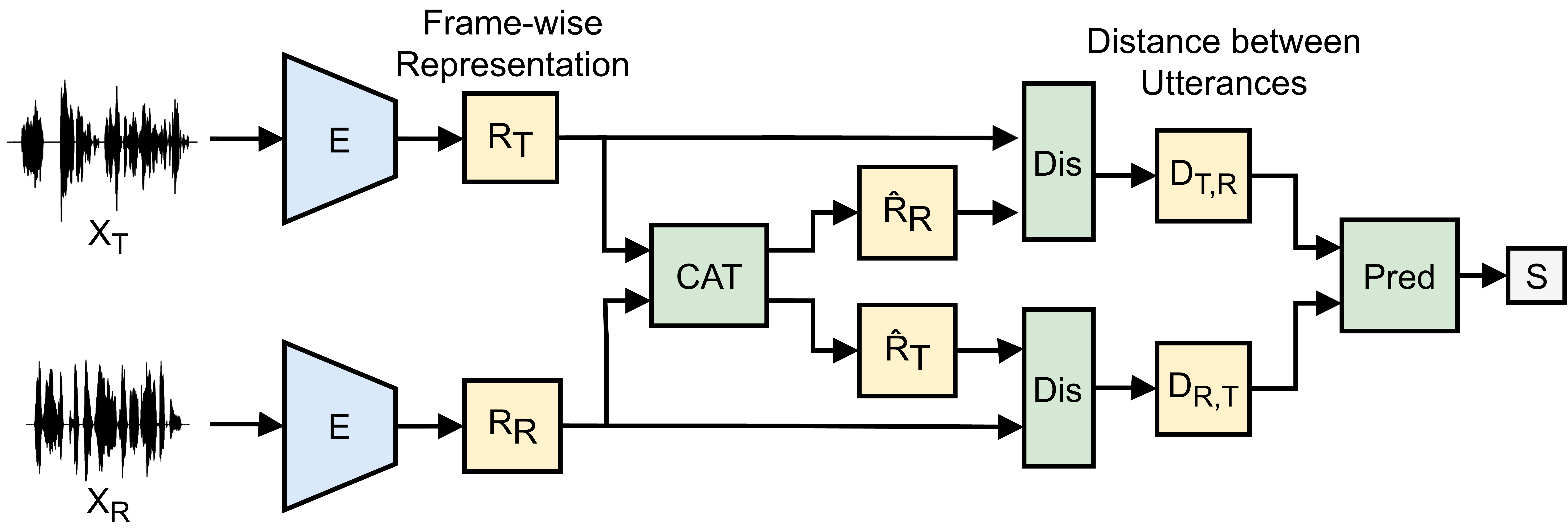}
        \vspace*{-0.8cm}
        \caption{}
        \label{fig:structure}
    \end{subfigure}
    \hfill
    \centering
    \begin{subfigure}[b]{0.27\textwidth}
        \centering
        \includegraphics[width=1\textwidth]{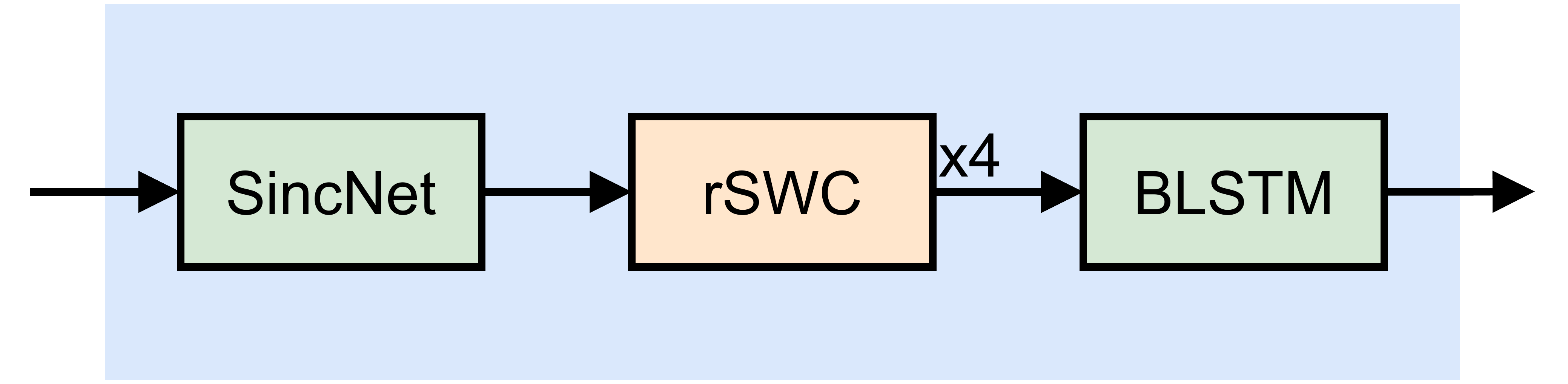}
        \vspace*{-0.5cm}
        \caption{}
        \label{fig:encoder}
    \end{subfigure}
    \hfill
    \begin{subfigure}[b]{0.4\textwidth}
        \centering
        \includegraphics[width=1\textwidth]{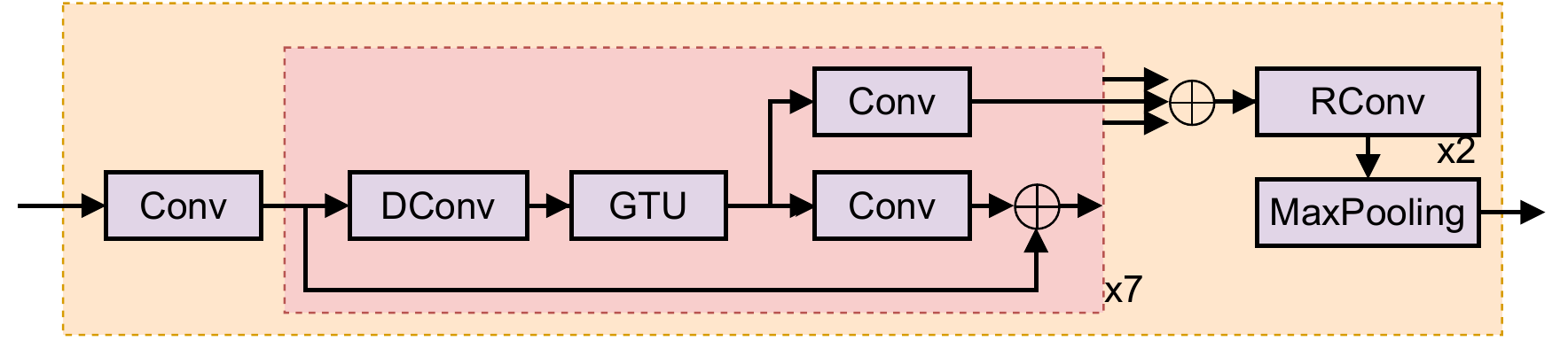}
        \vspace*{-0.5cm}
        \caption{}
        \label{fig:rswc}
    \end{subfigure}
    \vspace*{-0.2cm}
    \caption{(a) The architecture of SVSNet. E, CAT, Dis, and Pred blocks denote the Encoder, Co-attention, Distance, and Prediction modules. (b) The Encoder module. (c) The rSWC module. DConv is a dilated convolutional layer. RConv denotes a normal convolutional layer followed by a ReLU layer. The kernel size of the MaxPooling module is set to 3.}
    \label{fig:model}
    \hfill
    \vspace*{-0.5cm}
\end{figure}

Figure \ref{fig:model}(a) shows the SVSNet architecture. The encoder (E) module (shared by two inputs) encodes the waveforms of the test and reference utterances into frame-wise representations ($R_T$ and $R_R$). Unlike the attention module in \cite{zhang2019seq2seq}, which only aligns the test utterance with the enrollment utterance in one direction, to maintain the symmetry, the ``Co-attention'' module aligns the two representations in two directions. Then, two distances, namely $D_{T,R}$ (between $R_T$ and $\hat{R}_R$) and $D_{R,T}$ (between $R_R$ and $\hat{R}_T$), are computed by the ``Distance'' module and used to calculate the final similarity score by the ``Prediction'' module. We study two types of prediction modules: regression-based and classification-based. Their outputs are a continuous score and a score-level category, respectively. 

\vspace*{-0.05cm}
\subsection{Encoder}
Figure \ref{fig:model}(b) shows the architecture of the encoder in SVSNet. First, the input waveform is processed by SincNet, which contains $K$ learnable band-pass filters, to decompose the input signal to $K$ subband signals. 
The $K$ subband signals are then processed by four stacked residual-skipped-WaveNet convolution (rSWC) layers and a BLSTM layer.
Fig. \ref{fig:model}(c) shows the rSWC layer. The core of the rSWC layers is the convolutional layers with dilation sizes of (1, 2, 4, 8, 16, 32, and  64), followed by a gated tanh unit (GTU) \cite{gtu}. In addition, the maxpooling layer with a stride size of 3 is to downsample the feature sequence. As shown in Fig. \ref{fig:model}(a), given the test utterance $X_T$ and the reference utterance $X_R$, the encoder outputs $R_T$ and $R_R$, respectively.
\vspace*{-0.15cm}
\subsection{Co-attention Module}
The co-attention module is used to align the representation of the other input with that of one input:
\begin{equation}
\begin{aligned}
\hat{R}_R &= Attention(R_T, R_R, R_R), \\
\hat{R}_T &= Attention(R_R, R_T, R_T).
\end{aligned}
\end{equation}
Two pairs of aligned representation sequences are obtained, namely $(R_T, \hat{R_R})$ and $(R_R, \hat{R_T})$, which are then fed to the distance calculation module. We used the scaled dot-product attention mechanism \cite{vaswani2017attention} in this study.
\vspace*{-0.25cm}
\subsection{Distance Calculation and Prediction Modules}
We extend the attentive pooling used in SV \cite{Okabe2018} to our work. We average the representations of an utterance over time to obtain the utterance embedding and compute the 1-norm distance of each dimension of two means: 
\begin{equation}
\begin{aligned}
D_{T,R} &= ||Mean(R_T) - Mean(\hat{R_R})||_1, \\
D_{R,T} &= ||Mean(R_R) - Mean(\hat{R_T})||_1.
\end{aligned}
\label{eqn:utterance_dis}
\end{equation}
Then, the two distances are fed to the prediction module to obtain the similarity score:
\begin{equation}
\begin{aligned}
\hat{S}_T =& \sigma(f_{lin2}(\rho_{ReLU}(f_{lin1}(D_{T,R})))), \\
\hat{S}_R =& \sigma(f_{lin2}(\rho_{ReLU}(f_{lin1}(D_{R,T})))),
\end{aligned}
\end{equation}
where $\sigma(.)$ denotes an activation function, $f_{lin1}(.)$ and $f_{lin2}(.)$ denote two linear layers, and $\rho_{ReLU}(.)$ denotes a rectified linear unit (ReLU) activation function. The number of nodes of the final linear output layer is 1 for the regression model and 
4 for the classification model, i.e., $\hat{S}_T$, $\hat{S}_R$ $\in R^1$ for regression, and 
$R^4$ for classification. The activation function is an identity function for the regression model and a softmax function for the classification model. Finally, the prediction module obtains the final score by $\hat{S}$ = $(\hat{S}_T+\hat{S}_R)/2$.
\vspace*{-0.25cm}
\subsection{Model Training}
SVSNet is trained on a set of reference-test utterance pairs with corresponding human labeled similarity scores. We implemented two versions of SVSNet by using two types of prediction modules: regression and classification. The corresponding SVSNet models are termed SVSNet(R) and  SVSNet(C), respectively. Given the ground-truth similarity score $S$ and the predicted similarity score $\hat{S}$, the mean squared error (MSE) loss is used to train SVSNet(R), and the cross entropy (CE) loss is used to train SVSNet(C). 

\section{Experiments}

\subsection{Experimental Setup}
Since 2016, the VC challenge (VCC) has been held three times. The task is to modify an audio waveform so that it sounds as if it was from a specific target speaker other than the source speaker. In each challenge, a large-scale crowdsourced human perception evaluation was conducted to test the quality and similarity of the converted utterances. In this study, we focused on the similarity evaluation.

In VCC2018, there were 36 VC systems and two reference systems. Of the two reference systems, one took the source input as the output (used to evaluate the lower performance bound), and the other took the target output as the output (used to evaluate the upper performance bound). The similarity evaluation was conducted on 21,562 converted-natural utterance pairs, with two reference systems each accounting for 360 pairs, and the remaining systems each accounting for 570 to 599 pairs. Each pair was evaluated by 1 to 8 subjects, with a score ranging from 1 (same speaker) to 4 (different speakers). A total of 30,864 speaker similarity scores were obtained. The two reference systems were scored 614 and 614 times, and each of the remaining systems was scored 822 to 825 times. 
In addition, for each system, half of the pairs were converted-target pairs, and the other half of the pairs were converted-source pairs. The score of a system was the average score of its converted-target pairs. The detailed description of the corpus, listeners and evaluation methods can be found in \cite{lorenzo2018voice}. In this study, the dataset was divided into 24,864 pair-score samples for training and 6,000 pair-score for testing.


We used MOSNet \cite{lo2019mosnet} as the baseline. MOSNet was originally proposed for quality assessment, but a modified version was used for similarity assessment. Like SVSNet, the models via regression and classification are termed MOSNet(R) and MOSNet(C), respectively. Performance was evaluated in terms of accuracy (ACC), linear correlation coefficient (LCC) \cite{pearson1920notes}, Spearman’s rank correlation coefficient (SRCC) \cite{spearman1961proof}, and MSE at both utterance and system levels. 
The utterance-level evaluation was calculated from the predicted score and the ground-truth score for each pair of utterances. Note that if there was more than one score for a pair of utterances, the average value was used as the ground-truth score in testing. The system-level evaluation was calculated based on the average predicted score and the average ground-truth score for each system.
When treating similarity prediction as a classification task, the original labels were used as the ground-truth. When treating similarity prediction as a regression task and evaluating performance on the basis of ACC, the outputs of SVSNet(R) and MOSNet(R) were rounded and clipped to the nearest integer (i.e., 1, 2, 3, or 4).


Since two different sampling rates (22,050 and 16,000 Hz) were used in the VCC2018 dataset, we reduced the sampling rate of all utterances to 16,000 Hz.
For the encoder, the number of output channels of SincNet, the output size of the WaveNet convolutional layers, and the hidden size of BLSTM were 64, 64, and 256, respectively. The hidden size of the linear layers in the distance module was 128, and the output size was 1 and 4 for the regression output and classification output, respectively. 
We used the Adam optimizer to train the model, where the learning rate, $\beta_1$, and ${\beta_2}$ were 1e-4, 0.5, and 0.999, respectively. The batch size was set to 5. The model parameters were initialized by Xavier Uniform.


\vspace*{-0.25cm}
\subsection{Experimental Results on VCC2018}
First, we compare SVSNet with MOSNet. The results are shown in Table \ref{tab:4-class prediction}. 
From the table, several observations can be drawn. First, SVSNet consistently outperforms MOSNet in all metrics. Second, SVSNet performs better in regression mode than in classification mode, but MOSNet is difficult to determine which mode is better. Third, the high LCC (0.965 based on regression and 0.933 based on classification) and SRCC
(0.903 based on regression and 0.890 based on classification)
scores indicate that the predicted ranking of the 38
systems by SVSNet is close to that of human evaluation. 
\textcolor{red}{
}



Next, we further evaluate SVSNet(R) and SVSNet(C). We first study the effect of waveform processing. For a fair comparison, we replaced the waveform input and the SincNet in SVSNet(R) and SVSNet(C) with the 257-dimensional spectrogram used in MOSNet. The corresponding models are termed SVSNet(R)$_{spec}$ and SVSNet(C)$_{spec}$. From Table \ref{tab:ablation study}, we can see that SVSNet(R) is better than SVSNet(R)$_{spec}$ in all metrics, while SVSNet(C) is better than SVSNet(C)$_{spec}$ in all metrics except for the system-level MSE. The results confirm the advantage of using waveforms instead of spectrograms as input. 
Then, we study the effect of co-attention. In Table \ref{tab:ablation study}, SVSNet(R)$_{ss}$ and SVSNet(C)$_{ss}$ denote SVSNet with single-sided attention, which aligns the converted utterance with the natural utterance. SVSNet(R)$_{dtw}$ and SVSNet(C)$_{dtw}$ denote SVSNet with DTW, where the alignment is based on the dB-scale 2-norm distance between the spectrograms of two utterances. We can see that SVSNet(R)$_{ss}$ and SVSNet(C)$_{ss}$ are better than SVSNet(R)$_{dtw}$ and SVSNet(C)$_{dtw}$ in all metrics, respectively. Moreover, SVSNet(R) is better than SVSNet(R)$_{ss}$ in all metrics, while SVSNet(C) is better than SVSNet(C)$_{ss}$ in all metrics except for the system-level MSE. The results confirm the effectiveness of co-attention.

\begin{table}[!t]
\caption{Results of SVSNet and MOSNet on VCC2018.}
\label{tab:4-class prediction}
\fontsize{6.8}{7.8}\selectfont
\centering
\begin{tabular}{|l|p{0.53cm} p{0.53cm} p{0.53cm} p{0.53cm} |p{0.53cm} p{0.53cm} p{0.53cm}|}

\hline
\multirow{2}{*}{Method} & \multicolumn{4}{|c}{Utterance-level} & \multicolumn{3}{|c|}{System-level} \\
\cline{2-8}
 &  ACC & LCC & SRCC & MSE & LCC & SRCC & MSE \\
\hline
MOSNet(R)& 0.276 & \textbf{0.439} & 0.421 & \textbf{1.024} & \textbf{0.845} & \textbf{0.755} & 0.134 \\
MOSNet(C)& \textbf{0.432} & 0.426 & \textbf{0.434} & 1.463 & 0.796 & 0.665 & \textbf{0.087} \\
\hline
SVSNet(R)                & {0.438}     & \textbf{0.574}    & \textbf{0.572}    & \textbf{0.844}     & \textbf{0.965}    & \textbf{0.903}    & \textbf{0.044} \\
SVSNet(C)                & \textbf{0.482}     & {0.552}    & {0.554}    & {1.137}     & {0.933}    & {0.890}    & 0.057 \\
\hline
\end{tabular}
\end{table}

\begin{table}
\caption{Results of ablation study on SVSNet.}
\label{tab:ablation study}
\centering
\fontsize{6.8}{7.8}\selectfont
\begin{tabular}{|l|p{0.53cm}p{0.53cm}p{0.53cm}p{0.53cm}|p{0.53cm}p{0.53cm}p{0.53cm}|} 
\hline
\multicolumn{1}{|l|}{\multirow{2}{*}{Method}} & \multicolumn{4}{c|}{Utterance-level} & \multicolumn{3}{c|}{System-level} \\ 
\cline{2-8}
\multicolumn{1}{|l|}{} & ACC & LCC & SRCC & MSE & LCC & SRCC & MSE \\ 
\hline
SVSNet(R) & \textbf{0.438}     & \textbf{0.574}    & \textbf{0.572}    & \textbf{0.844}     & \textbf{0.965}    & \textbf{0.903}    & \textbf{0.044} \\
SVSNet(R)$_{spec}$ & 0.417  & 0.545 & 0.544 & 0.879 & 0.929 & 0.857 & 0.067 \\
SVSNet(R)$_{ss}$ & 0.404 & 0.550 & 0.552 & 0.874 & 0.929 & 0.811 & 0.092 \\
SVSNet(R)$_{dtw}$ & 0.353  & 0.534 & 0.527 & 0.906  & 0.873 & 0.809 & 0.093 \\
\hline
SVSNet(C) & 0.482     & 0.552    & 0.554    & 1.137     & 0.933    & 0.890    & 0.057 \\
SVSNet(C)$_{spec}$ & 0.459  & 0.489 & 0.492 & 1.211  & 0.904 & 0.832 & 0.053 \\
SVSNet(C)$_{ss}$ & 0.481 & 0.531 & 0.534 & 1.164 & 0.932 & 0.822 & 0.044 \\
SVSNet(C)$_{dtw}$ & 0.462 & 0.512 & 0.517 & 1.173  & 0.886 & 0.770 & 0.071 \\
\hline
\end{tabular}
    \vspace*{-0.25cm}
\end{table}
     \vspace*{-0.35cm}

\begin{figure}
    \vspace*{-0.3cm}
    \centering
    \begin{subfigure}[b]{0.22\textwidth}
    \includegraphics[width=\textwidth]{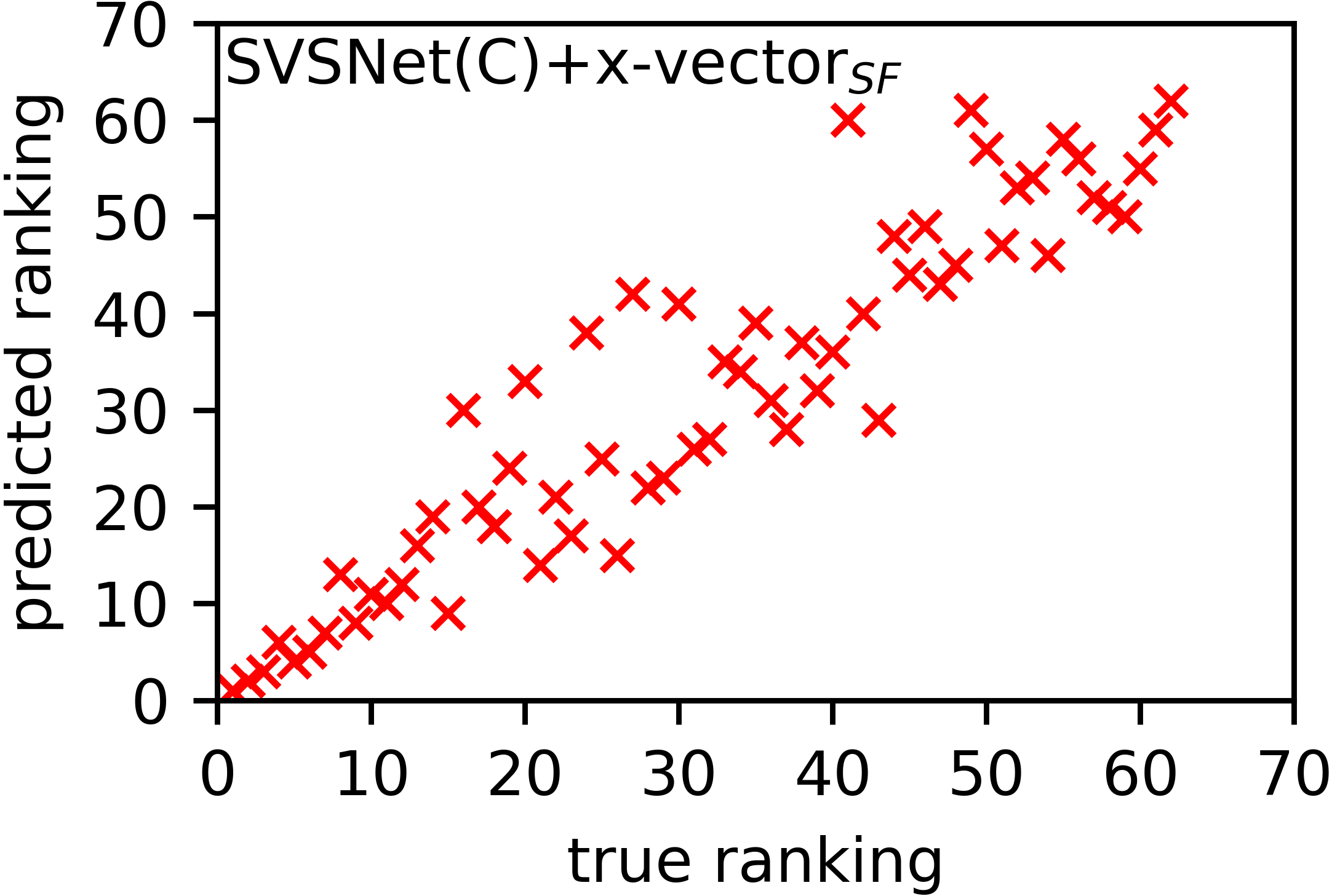}
    \end{subfigure}
    \hfill
    \begin{subfigure}[b]{0.22\textwidth}
    \includegraphics[width=\textwidth]{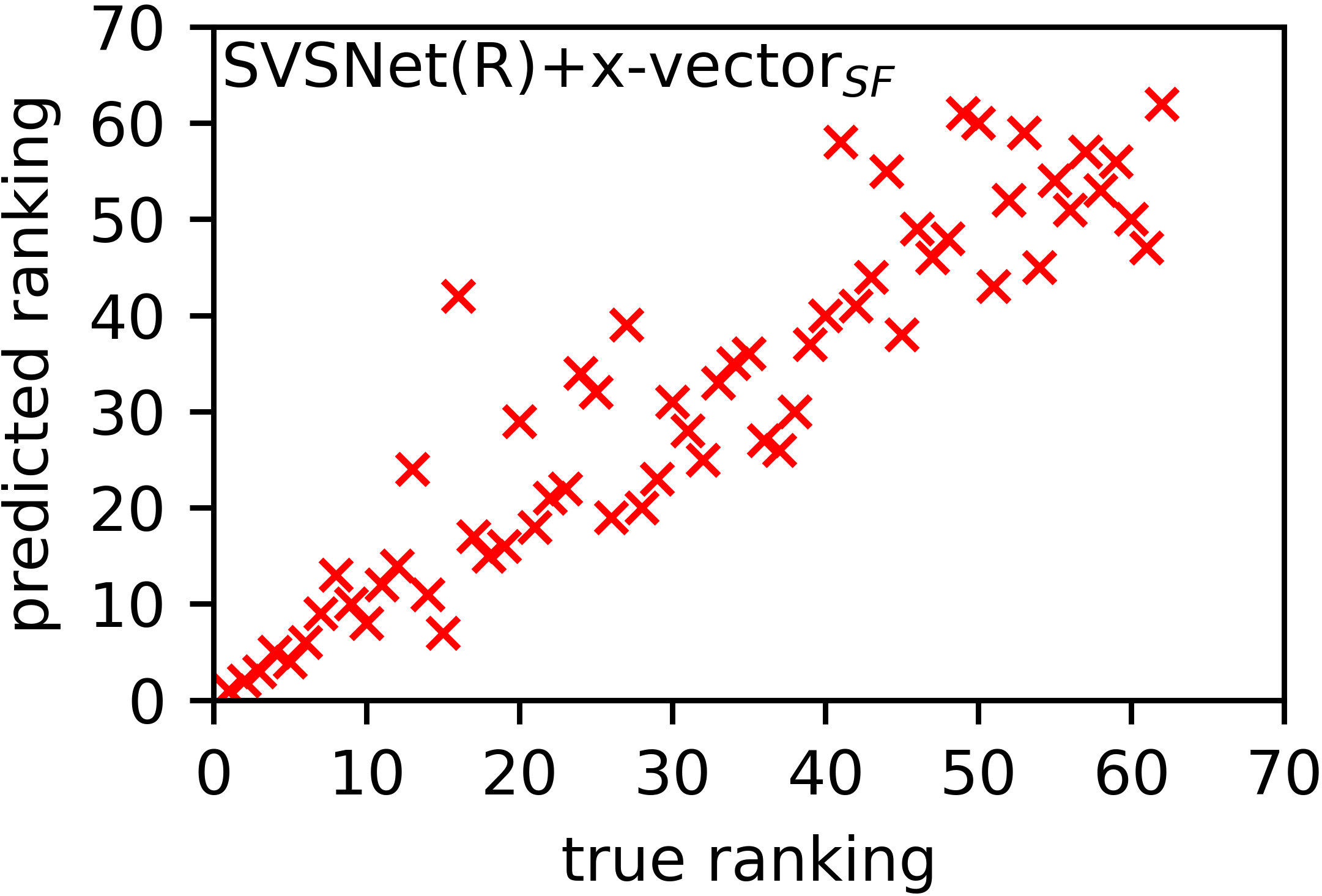}
    \end{subfigure}
    \caption{Predicted system rankings of the score fusion models on VCC2020.}
    \label{fig:scatter plot}
    \vspace*{-0.2cm}
\end{figure}

\subsection{Experiment results on VCC2020}
Voice Conversion Challenge 2020 (VCC2020), the next edition of VCC2018, includes two tasks, namely intra-language VC and cross-language VC. The intra-language task considered 16 source-target speaker pairs, and the cross-language task considered 24 source-target speaker pairs. Each source-target speaker pair contained 5 converted-target utterance pairs, and each converted-target utterance pair was evaluated by 12 subjects (for intra-language) and 8 subjects (for cross-language). Therefore, there were 960 evaluation scores per system (16$\times$5$\times$12 for intra-language and 24$\times$5$\times$8 for cross-language).
There were 31 VC systems for the intra-language task, 28 VC systems for the cross-language task, and three reference systems for evaluating the lower and upper performance bounds.
To study the impact of corpus mismatch, we tested the models trained on the VCC2018 training set on the complete VCC2020 dataset to perform system-level evaluation. Since most systems used conventional vocoders in VCC2018 and neural vocoders in VCC2020, the corpus mismatch is significant. It is also worth mentioning that SVSNet(R)$_{dtw}$ and SVSNet(C)$_{dtw}$ are not applicable because the two utterances to be compared here have different content.

The results are shown in Table \ref{tab:vcc20}. We can see that the scores of both SVSNet and MOSNet are lower than those reported earlier due to corpus mismatch, while SVSNet still outperforms MOSNet. Following Das et al. \cite{Das2020}, we tested the performance with another prediction model formed by a cosine similarity measure based on 128-dimensional linear discriminant analysis (LDA) reduced x-vectors. The similarity scores were linearly mapped to [1, 4]. The results show that with an extra and massive dataset for pretraining, the x-vector model outperforms both SVSNet and MOSNet on LCC and SRCC. We also constructed fusion models to utilize the information of x-vector into SVSNet. 
For the score fusion models (SVSNet(R)+x-vector$_{SF}$ and SVSNet(C)+x-vector$_{SF}$), the fusion score was the weighted average of the scores of the SVSNet and x-vector models at a ratio of 3:7. The weight was determined based on the test set of VCC2018. For the feature fusion models (SVSNet(R)+x-vector$_{
FF}$ and SVSNet(C)+x-vector$_{FF}$), the 1-norm distance of each dimension between two x-vectors was concatenated to $D_{T,R}$ and $D_{R,T}$ in Eq. \ref{eqn:utterance_dis}. From Table \ref{tab:vcc20}, we can see that all fusion models yield improvements over the SVSNet and x-vector models.
Since the score fusion models achieve the best SRCC values in Table \ref{tab:vcc20}, we compare their predicted system rankings with the true ranking in Fig. \ref{fig:scatter plot}. Obviously, both models achieve fairly good prediction performance on VCC2020, although there is still room for further improvement.

\begin{table}
    \caption{Results of different models on VCC2020.}
    \label{tab:vcc20}
\centering
\fontsize{6.8}{7.8}\selectfont
\begin{tabular}{|l|lll|} 
\hline
\multirow{2}{*}{Method} & \multicolumn{3}{c|}{System-level} \\ 
\cline{2-4}
 & LCC & SRCC & MSE \\ 
\hline
MOSNet(R) & 0.630 & 0.504 & 2.964 \\
MOSNet(C) & 0.723 & 0.736 & 0.211 \\ 
\hline
x-vector & 0.844 & 0.911 & 1.213 \\
\hline
SVSNet(R) & 0.819 & 0.841 & 0.361 \\
SVSNet(C) &  0.775 & 0.765 & 0.244 \\ 
\hline
SVSNet(R)+x-vector$_{SF}$ & 0.872 & 0.926 & 0.867 \\
SVSNet(C)+x-vector$_{SF}$ & 0.877 & \textbf{0.936} & 0.711 \\ 
\hline
SVSNet(R)+x-vector$_{FF}$ & 0.863 & 0.921 & \textbf{0.141} \\
SVSNet(C)+x-vector$_{FF}$ & \textbf{0.885} & 0.926 & 0.373 \\ 
\hline
\end{tabular}
    \vspace*{-0.2cm}
\end{table}


\section{Conclusions}
In this paper, we have proposed SVSNet, an end-to-end neural similarity assessment model. The results of experiments on the large-scale human perception evaluation in VCC2018 and VCC2020 show that SVSNet, benefiting from the SincNet and the residual-skipped-WaveNet architecture, performs better than the previous model MOSNet in terms of LCC, SRCC, and MSE. It is also found that directly using the waveform as input without discarding the phase information will increase the prediction ability of our model. In the future, we plan to consider the theory of human perception to design a perception-based objective function to build a more robust quality and similarity prediction model. We will also explore the use of SVSNet in the training objective to guide VC models to generate utterances that are highly similar to the target speaker's speech.

\bibliographystyle{ieeetr}
\bibliography{ref.bib}
%








\end{document}